%% file: main.tex
 \documentclass[reprint]{JASA}

\begin{document}

\title[JASA/Sample JASA Article]{Pointwise Encoding Time reduction with Radial Acquisition (PETRA) to pseudo-CT mapping for precision Focused Transcranial Ultrasound Applications}

\author{Maria Miscouridou}
\email{maria.miscouridou.16@ucl.ac.uk}
\affiliation{Department of Medical Physics and Biomedical Engineering, University College London, London, WC1E 6BT, UK}

\author{Alisa A. Krokhmal\footnote{current address: Lomonosov Moscow State University}}
\affiliation{Department of Medical Physics and Biomedical Engineering, University College London, London, WC1E 6BT, UK}

\author{Aaron T. Hess}
\author{Jon Campbell}
\author{Charlotte J. Stagg}
\affiliation{Nuffield Department of Clinical Neurosciences, FMRIB, Wellcome Centre for Integrative Neuroimaging, University of Oxford, Oxford, OX3 9DU, UK}

\author{Eleanor Martin}
\author{Bradley E. Treeby}
\affiliation{Department of Medical Physics and Biomedical Engineering, University College London, London, WC1E 6BT, UK}

\date{\today} 

\newcommand{\changed}[1]{\textcolor{orange}{#1}}
\newcommand{\review}[1]{\textcolor{red}{#1}}

\input{0Abstract}     

\maketitle

\input{1Introduction}

\input{3InVivo}
\input{4ExVivoMethods}

\input{4ExVivoResults}
\input{5Conclusions}



\begin{acknowledgments}
The following article has been submitted to JASA. After it is published, it will be found at https://pubs.aip.org/asa/jasa.
This research was supported by the Engineering and Physical Sciences Research Council, UK (EP/S026371/1, EP/S021612/1),
as well as the UKRI Future Leaders Fellowship [MR/T019166/1].
The Wellcome Centre for Integrative Neuroimaging is supported by core funding from the Wellcome Trust (203139/Z/16/Z). CJS holds a Senior Research Fellowship, funded by the Wellcome Trust (224430/Z/21/Z). 
This work was also supported by the Ministry of Education, Youth and Sports of the Czech Republic through the e-INFRA CZ (ID:90254). 
\end{acknowledgments}



\docsection{Author Declarations}
The authors have no conflicts to disclose.

\docsection{Data availability}
The pseudo-CT generation code is available as a GitHub repository \cite{petra_repo}. The MR and CT data that support the findings of this study can be available from the corresponding author upon reasonable request. The measurement data that support the findings of this study are openly available at the UCL Research Data Repository at \url{https://doi.org/10.5522/04/28025240.v2} \cite{meas_data}.


\bibliography{bibliography}

\end{document}

%% file: 0Abstract.tex
\begin{abstract}

Ultrasound simulations used for treatment planning in transcranial ultrasound stimulation(TUS) currently depend on converting computed tomography(CT) images to acoustic properties.
Aiming to reduce subject exposure to ionising radiation, recent literature focused on generating pseudo-CT images from different magnetic resonance(MR) imaging sequences.
Here, the PETRA (pointwise encoding time reduction with radial acquisition) sequence was used because of its stronger ability to image bones.
A dataset of paired CT and PETRA scans was acquired separately for human subjects and ex-vivo skulls. Principal component analysis performed on all bone voxels uncovered the affine relationship converting PETRA images to pseudo-CTs.
Acoustic simulations based on CTs and pseudo-CTs were performed in \texttt{k-Plan} and k-wave.
For in-vivo data, the mean absolute errors in focal position, peak pressure and focal volume were 0.48±0.25mm, 5.1±4.0\% and 5.8±4.4\%.
The similarity of acoustic fields from ex-vivo CTs/pseudo-CTs was compared with experimental data. The corresponding errors were 1.5±1.2 mm, 17.1±14.6\% and 32.5±22.5\% for CT and 1.6±1.4 mm, 18.2±17.1\% and 19.2±15.9\% for pseudo-CT. 
The small errors for in-vivo data and the similarity of the experimental errors for both methods validate the use of PETRA-derived pseudo-CT as an alternative to CT, ensuring precise acoustic field predictions. The conversion tool is available at \url{https://github.com/ucl-bug/petra-to-ct}.

\end{abstract}

%% file: 1Introduction.tex
\section{Introduction}\label{sec:introduction}
Transcranial ultrasound stimulation (TUS) is an emerging therapeutic technique that leverages ultrasound waves to modulate brain activity \citep{darmani2022noninvasive}. A factor that needs to be taken into account is the aberration caused to the waves by the skull. For this reason ultrasound simulations are currently used to aid with target guidance and dosimetry. Currently, simulations are paired with x-ray computed tomography (CT) images of individual subjects to generate personalised maps of the acoustic properties and predict how the ultrasound waves propagate through the bones \citep{aubry2003experimental, marquet2009non}. Although CT images are the gold standard for bone imaging, they are not always available and they expose patients to ionising radiation, which motivates us to explore methods that avoid CT scans. Our approach is to translate from magnetic resonance (MR) images to CT images. So far, classical and machine learning techniques have been used to generate pseudo-CTs (pCT) from a range of MR sequences, including T1-weighted, UTE (ultrashort echo time) and ZTE (zero echo time) \citep{miscouridou2022classical, su2020transcranial, yaakub2023pseudoCT}.

The existing literature on translating MR to CT can be divided into three different approaches. 
Before the use of machine learning, classical methods have been used to translate MR to pCT.
First, piece-wise constant methods segment the MR into brain, skull and skin/background areas and use predefined values in Hounsfield Units for each area to generate a pCT \citep{wintermark2014t1}. The segmentation process usually works quite well and the interfaces between different types of tissue can be identified accurately. However, one value in each segment results in the homogenisation of the internal structures and does not represent the complex microstructure. In particular, trabecular and cortical bone have different 
acoustic properties and treating them as homogeneous introduces errors in the simulation.

Second, deep learning (DL) techniques have been shown to accurately map different types of MR images to pCTs. 
A more extensive review of such methods can be found in previous work \citep{miscouridou2022classical}. T1-weighted MR images are most commonly used as they are the most prevalent in the medical field.
These methods can provide more accurate predictions than classical methods for the smaller and more heterogeneous structures of the bone. However, they suffer from poor generalisation to new data, a common drawback of DL methods.
Third, specialised MR sequences such as ZTE or UTE can be used to directly map to CT. These methods involve segmenting the MR and using direct mapping from them to pseudo-CT. Such specialised imaging protocols can capture bone signals with greater success compared to T1-weighted MRIs. Therefore, it is possible to infer an affine relationship between MR values and CT values in the bone. These can then be used to generate pCTs directly from MRIs \citep{marquet2009non}.

Most conventional MRI techniques (such as T1-weighted) struggle to visualise structures such as bone and tendons because of their rapid signal decay. Specialised MRI techniques are used to overcome the challenges associated with imaging these tissues. The most common sequences that have demonstrated bone imaging are UTE \cite{gatehouse2003magnetic} and ZTE \citep{weiger2012mri} because of their short echo times. 
Another sequence that can be optimised to capture bone signals is PETRA (pointwise encoding time reduction with radial acquisition) \citep{grodzki2021ultrashort}.


A low flip angle can be chosen to minimise T1 contrast in PETRA scans, affecting the appearance of tissues such as the cerebrospinal fluid (CSF). In conventional T1-weighted images, CSF appears dark due to its longer T1 relaxation time. By reducing T1 contrast, the CSF appears less dark, improving the visibility of bones. It allows for cortical bone and cortical bone water imaging \citep{li2017selective}, resulting in better contrast between bone and surrounding tissues. Recently, PETRA has proved a valuable tool in musculoskeletal and orthopaedic imaging \citep{kim2018clinical}, since it is designed to provide high signal from tissues with very short T2 times, such as bone and tendons. 
Additionally, PETRA has been shown to reduce metal artefacts in dental imaging, although it can lead to poorer image quality \citep{hilgenfeld2017petra}.



While all MRI sequences sample data in k-space, PETRA's uniqueness lies in how it optimises the acquisition and reconstruction process. PETRA differs from ZTE in how it starts signal acquisition; while ZTE achieves a true zero echo time (TE = 0), it has a gap in the centre of k-space due to receiver dead time, leading to image blurring. PETRA solves this gap issue by sampling the center of k-space with a Cartesian approach. This hybrid Cartesian - radial encoding of PETRA makes it also more efficient than UTE in bone imaging. PETRA encodes the very center of k-space with Cartesian encoding which prevents loss of short-T2* signals. UTE uses fully radial acquisition and is therefore less efficient in capturing short-T2 tissues, such as cortical bone and myelin.
Additionally, the sampling strategy in PETRA scans can capture essential low-frequency components more efficiently than ZTE. This leads to better contrast and detail of the bony structures within the skull and minimisation of truncation artefacts such as Gibbs ringing.
The radial acquisition pattern used reduces sensitivity to motion artefacts which also makes it a better choice for imaging moving organs like the heart and lungs \citep{nozawa2019imaging}. Other than its advantages in data handling, PETRA is available in Siemens scanners, which are widely used in clinical settings.


These features collectively render PETRA a robust option for skull imaging, providing a well-balanced presentation of detail, contrast, and safety.
Although the advantages of PETRA sequences mentioned above suggest it is a strong candidate for pCT generation, very little work has been done in the area \citep{ghose2017substitute} and none of it involves humans or head scans.
This motivates us to investigate the accuracy of PETRA to CT conversion, specifically for human head scans, and use this conversion to
generate personalised pCTs from both in-vivo and ex-vivo data. 
Our objective is to develop an accurate, validated and accessible skull mapping method that eliminates the need for CT and can be used in acoustic simulations.

%% file: 3InVivo.tex
\section{Implementation and Numerical Evaluation on Humans}\label{sec:invivo}

\subsection{PETRA to pseudo-CT}\label{sec:pctMethods}

This section uses a method previously described by \citep{miscouridou2022classical} which converts MRI scans into pseudo-CT images, this time applying it to PETRA images. It outlines the data acquisition procedure and all required processing steps. Additionally, it provides a direct PETRA to CT mapping method that serves as a practical guide, enabling users to both reproduce these results and generate their own pseudo-CT images.

\subsubsection{Data acquisition}
The dataset acquired for this study consisted of paired PETRA and CT scans of seven subjects. The CT images are low-dose CTs and the typical effective dose for them is $< 0.2$ mSv. Note that during MR acquisition a 64-channel Head/Neck coil was used and 3D distortion correction was applied to the data post acquisition. For a user that wants to reproduce these results and generate their own pCTs, we propose the set of the following acquisition parameters for CT and MR scans in Table~\ref{tab:acquisition_params}. The full set of parameters and the equivalent EXAR file can be found in the repository \citep{petra_repo}.

\begin{table}[h]
\begin{tabular}{c|c}
\toprule
\textbf{MR} & \textbf{CT} \\
\midrule
\multicolumn{2}{c}{System} \\ 
\midrule
Siemens Magnetom Prisma & GE Revolution \\
Magnetic field strength: 3T & \\
Software: Syngo MR E11 & low-dose scans \\
\midrule
\multicolumn{2}{c}{Parameters} \\ 
\midrule
TR 1 - 3.61 ms & Convolution kernel: BONEPLUS\\
TE – 0.07 ms & Tube current: 63 mA \\
Flip angle – 1 \degree & kVp: 80 \\
Slices per slab – 320 & Pixel spacing: 0.44 mm \\
Slice thickness – 0.75 mm & Slice thickness: 0.625 mm\\
\bottomrule
\end{tabular}
\caption{Acquisition parameters used for MR and CT imaging.}
\label{tab:acquisition_params}
\end{table}


\begin{figure*}[t]
    \centering
    \includegraphics[width=\textwidth]{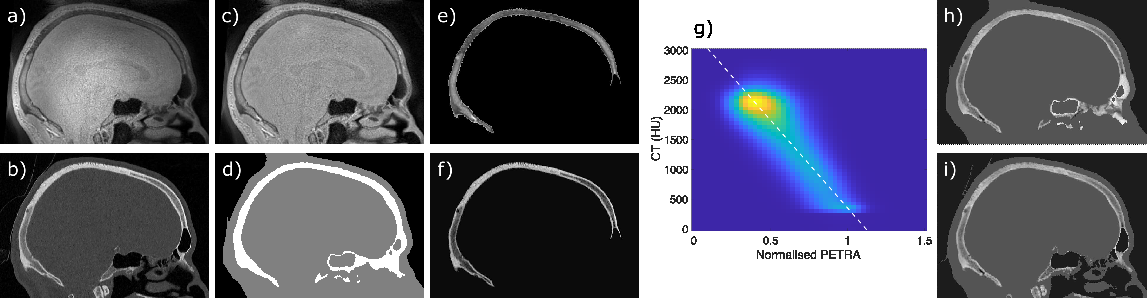}
    \caption{Middle sagittal slices of the following images: a) PETRA registered to CT b) CT c) debiased PETRA d) head, bone and air masks generated with SPM e) skull only in PETRA f) skull only in CT h) pseudo-CT with SPM-derived masks i) pseudo-CT with CT-derived masks g) correlation plot of normalised PETRA and CT.}
    \label{fig:pipeline}
\end{figure*}

\subsubsection{Image preprocessing}\label{sec:preprocessing}

The images were processed as follows. First, the PETRA scans were registered to the CT scans using FSL \cite{jenkinson2001global, jenkinson2002improved}, a widely used tool which allows using the normalised mutual information cost function and 12 degrees of freedom. As part of the registration, the MR images were resampled to match the resolution and field-of-view of the CT images. The registered PETRA scans were then debiased using N4ITK \cite{tustison2010improved}. Removing the bias-field helps to reduce image intensity non-uniformities resulting from transmit RF inhomogeneities and receive coil sensitivities. Then, histogram normalisation was applied to each PETRA image separately to shift the soft-tissue peak to 1. This ensures the intensity distributions are consistent across images and ensures the method is applicable to external data. 

Lastly, in order to segment the different tissues of the head, masks were generated using MR and CT separately. Firstly, masks were generated using the CT scan of each subject since the tissue boundaries are clearer and the segmentation is more accurate. Bone, brain and air masks were created by performing a series of thresholding and morphological operations on the CT using  \citep{kplan}.
A skull mask was derived by removing bones voxels outside the neurocranium (e.g teeth, spine) using MATLAB.
Additionally, head and skull masks were created using the PETRA images using SPM12 \cite{spm}. followed by morphological operations in MATLAB. This is because the skull mask tool in k-Plan uses CT and is not effective with MR. The exact use of each mask is explained in the relevant sections following. Examples of the raw and processed PETRA and CT images are shown in Fig.~\ref{fig:example_pcts}.

\subsubsection{MR to CT Mapping} 
To obtain the mapping from PETRA to CT HU values, only voxels in the skull mask should be considered and bones outside the neurocranium should be ignored, shown in Fig.~\ref{fig:pipeline} e) and f). For this purpose, the skull mask derived from the CT was used, described in the section above. The mapping was extracted by taking the first principal component of the density plot of normalised PETRA values and CT values. The resulting affine relationship is \begin{equation}\label{eq:petratoctlinear}
\text{CT} = -2940.2 \cdot \text{PETRA} + 3291.1.
\end{equation} as shown by the line in Fig.~\ref{fig:pipeline} g).


\subsubsection{Pseudo-CT generation}
\begin{figure*}[]
    \centering
    \includegraphics[]
    {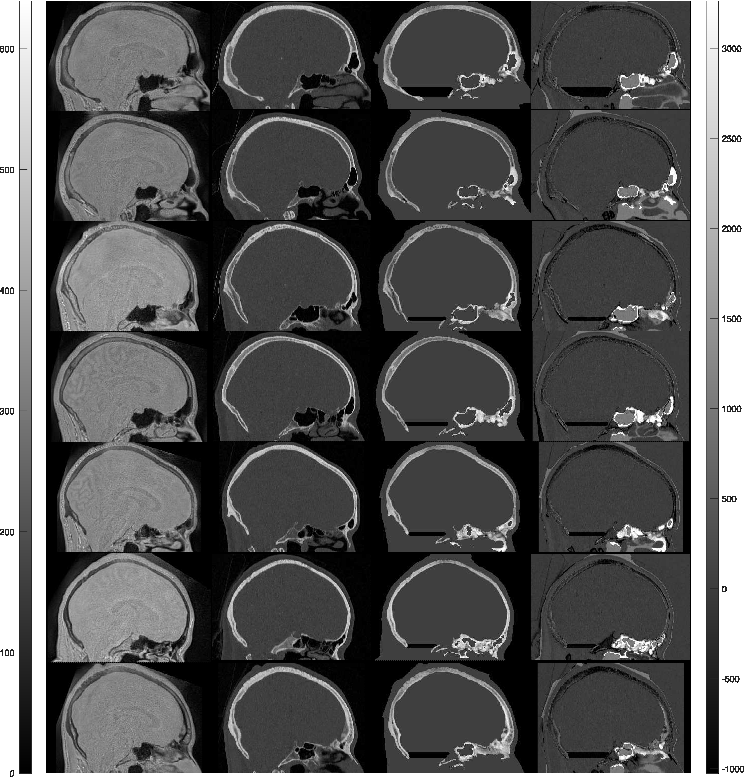}
    \caption{Middle sagittal slice for PETRA, CT, pCT and error map (columns) for all 7 subjects (rows).}
    \label{fig:example_pcts}
\end{figure*}

To generate the pCTs, masks produced using SPM were used. This is because access to a CT scan would not be available to a typical user and masks would therefore need to be derived from the MR images. Voxels in the head mask were assigned to 42 HU, voxels in the skull mask were converted with the affine mapping from Eq.~\ref{eq:petratoctlinear}. All other voxels in the background/air were assigned to -1000 HU. Following this methodology, pseudo-CTs were generated in 3D for the 7 subjects in the dataset and they are displayed in \ref{fig:example_pcts}.

\subsubsection{Image metrics and results}\label{sec:pctevaluation}

The pCTs were compared against the ground truth CT to evaluate their difference voxel-wise in Hounsfield Units. To evaluate them, the mean absolute error (MAE) and the root mean squared error (RMSE) were chosen as metrics. Comparing the CT and pCT across subjects, the average MAE was 203 $\pm$ 16 HU in the head mask, 347 $\pm$ 25 HU in the skull mask and 64 $\pm$ 1 HU in the brain mask. Results for individual subjects are given in Table~\ref{tab:mae_rmse}. For reference, the average value of the CT was $37 \pm 82$ HU in the brain and $1574 \pm 590 $ HU in the skull. The masks used in the evaluation were the ones derived from CTs as they segment the different regions more accurately.

\begin{table}[h]
\centering
\begin{tabular}{c|cc|cc}
\toprule
 & \multicolumn{2}{c|}{\textbf{MAE} (HU)} & \multicolumn{2}{c}{\textbf{RMSE} (HU)}  \\
\toprule
\diagbox[height=1.5\line]{\textbf{Subject}}{\textbf{Mask}} & \textbf{head} & \textbf{skull} & \textbf{head} & \textbf{skull} \\
\midrule
\textbf{S1} & 222 & 314 & 486 & 395 \\
\textbf{S2} & 230 & 317 & 506 & 405 \\
\textbf{S3} & 189 & 355 & 420 & 444 \\
\textbf{S4} & 200 & 332 & 453 & 421 \\
\textbf{S5} & 193 & 369 & 451 & 460 \\
\textbf{S6} & 191 & 374 & 428 & 502 \\
\textbf{S7} & 198 & 365 & 452 & 455 \\
\midrule
\textbf{mean} & 203$\pm$16 & 347$\pm$25 & 457$\pm$30 & 440$\pm$37  \\
\bottomrule
\end{tabular}
\caption{Mean absolute error (MAE) and root mean squared error (RMSE) for the 3D pCT images generated compared against the ground truth CT images, separately for each subject.}
\label{tab:mae_rmse}
\end{table}

\section{Acoustic simulations}

\subsection{Simulation setup}\label{sec:acoustic_simulation}

Apart from standard image comparison metrics, the pCTs were compared to CTs by evaluating the differences in their effect on simulated ultrasound fields. For this purpose, 3D acoustic simulations were performed in \texttt{k-Plan} \citep{kplan} for pCTs and ground truth CTs. K-Plan performs linear acoustic simulations using k-Wave \cite{treeby2010k}.
Within k-Plan, the image is segmented into skull, soft tissue (brain and skin) and background regions with thresholding and morphological operations. 
First, the Hounsfield units of the CT image are converted to mass density. To calibrate this conversion, a CT image of a CIRS Model 062M Electron Density Phantom was acquired using the same acquisition settings, and the extracted curve is used within the skull. 
The sound speed $c$ of the skull is calculated from the density values $\rho$ using the relationship of $c = 1.33 \rho + 166.7$ \cite{marquet2009non}. The attenuation coefficient used for the skull is $13.3\cdot f$ dB/cm, where $f$ is frequency in MHz \cite{pinton2012attenuation}. 
All tissue in the head except the skull is considered soft tissue and has density 1045 kg/m$^3$, sound speed 1550 m/s and attenuation coefficient 0.059 dB/m.  . 
The background region is assigned the properties of water at 30$\degree$C with density 996 kg/m$^3$, sound speed 1509 m/s and attenuation coefficient 0.00217 dB/m. 
Simulations were run at 6 points per wavelength and 2 grid traversals.

The transducer used was NEUROFUS-CTX-500-4, which is a 4-element annular array with ROC 63.2 mm. It was driven at 500 kHz and was targeted at the occipital pole of the primary visual cortex and the hand knob of the primary motor cortex in both hemispheres, resulting in 4 targets per subject. These targets were selected manually from a T1-weighted scan of each subject, which was co-registered to the CT scans with the same method used for the PETRA scans. All the other parameters used (source geometry, source position, target position, sonication frequency, sonication pulse parameter) were identical in the CT and pCT setup so that the results were comparable.

\subsection{Acoustic metrics}\label{sec:acoustic_metrics}

The results of the simulations were compared per subject to assess if the CT and pCT could produce similar acoustic fields.
The evaluation assessed the accuracy of the spatial peak pressure amplitude, the volume of the -6 dB focal zone (which encompasses the region where the pressure is within -6 dB of the peak pressure), and the focal position. 
These three metrics which were established in \cite{aubry2022benchmark} were calculated as below, with code available from \cite{intercomparisoncode}:

\begin{itemize}
    \item Difference in focal position calculated as the centre of mass of the focal volume, in mm:
        \begin{equation}
            r = ||x_{2} - x_{1}, y_{2} - y_{1}, z_{2} - z_{1}||, 
        \end{equation}
    \item Difference in spatial peak positive pressure within the focal volume, expressed as percentage:
        \begin{equation}
            \varepsilon_p = 100 * (p_{2} - p_{1}) / p_{1}, 
        \end{equation}    
    \item Difference in focal volume, expressed as percentage:
        \begin{equation}
            \varepsilon_V = 100 * (V_{2} - V_{1})/V_{1},
        \end{equation}
\end{itemize}
where subscripts 1 and 2 correspond to CT and pCT derived simulations respectively.

\subsection{Acoustic evaluation and results}
\begin{figure*}
    \centering
    \begin{subfigure}{0.49\linewidth}
        \includegraphics[width=\linewidth]{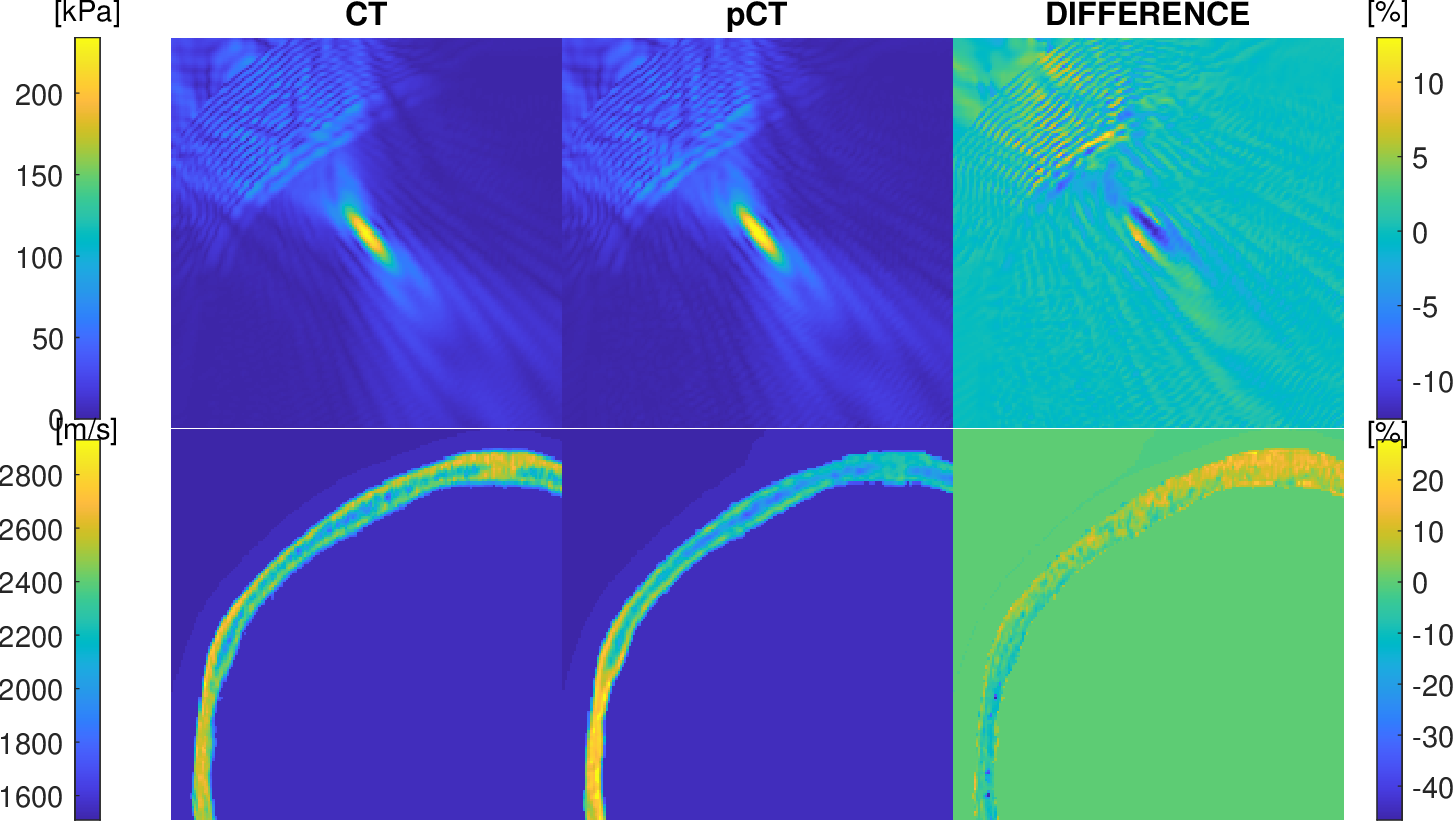}
	    \caption{left motor cortex}
	\end{subfigure}
	\begin{subfigure}{0.49\linewidth}
        \includegraphics[width=\linewidth]{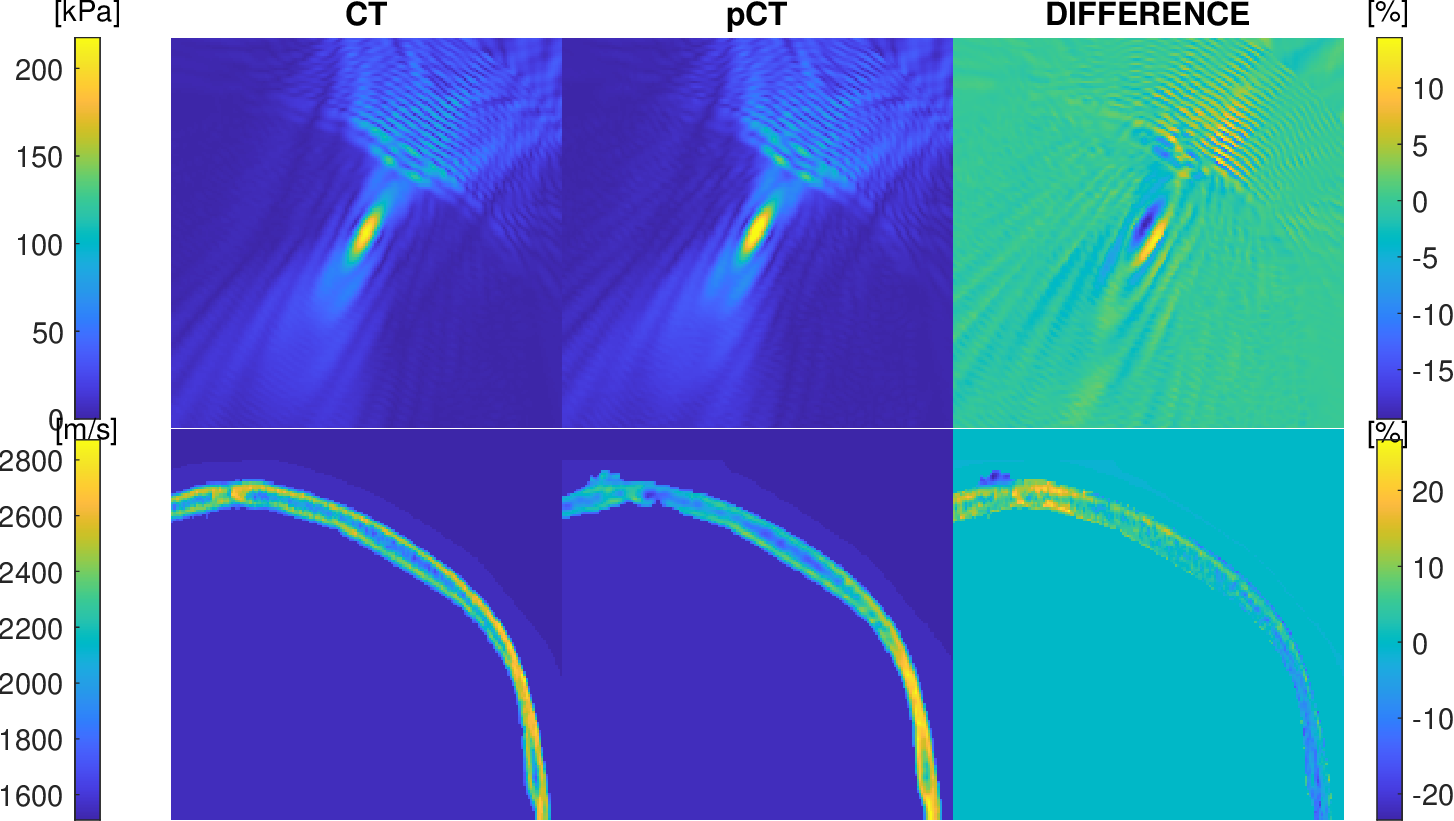}
	    \caption{right motor cortex}
    \end{subfigure}
    \vfill
    \begin{subfigure}{0.49\linewidth}
        \includegraphics[width=\linewidth]{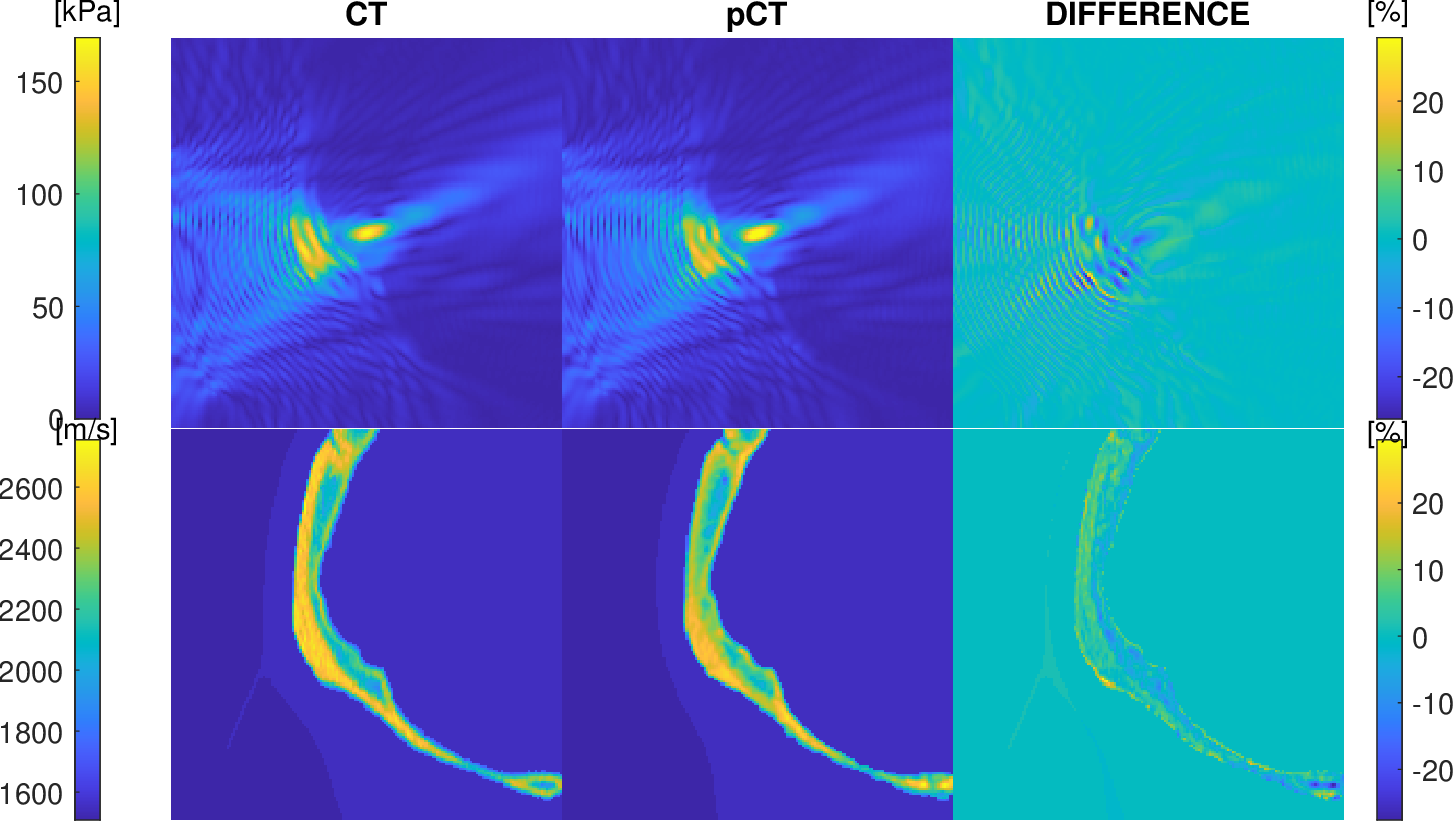}
	    \caption{left visual cortex}
	\end{subfigure}
    \begin{subfigure}{0.49\linewidth}
        \includegraphics[width=\linewidth]{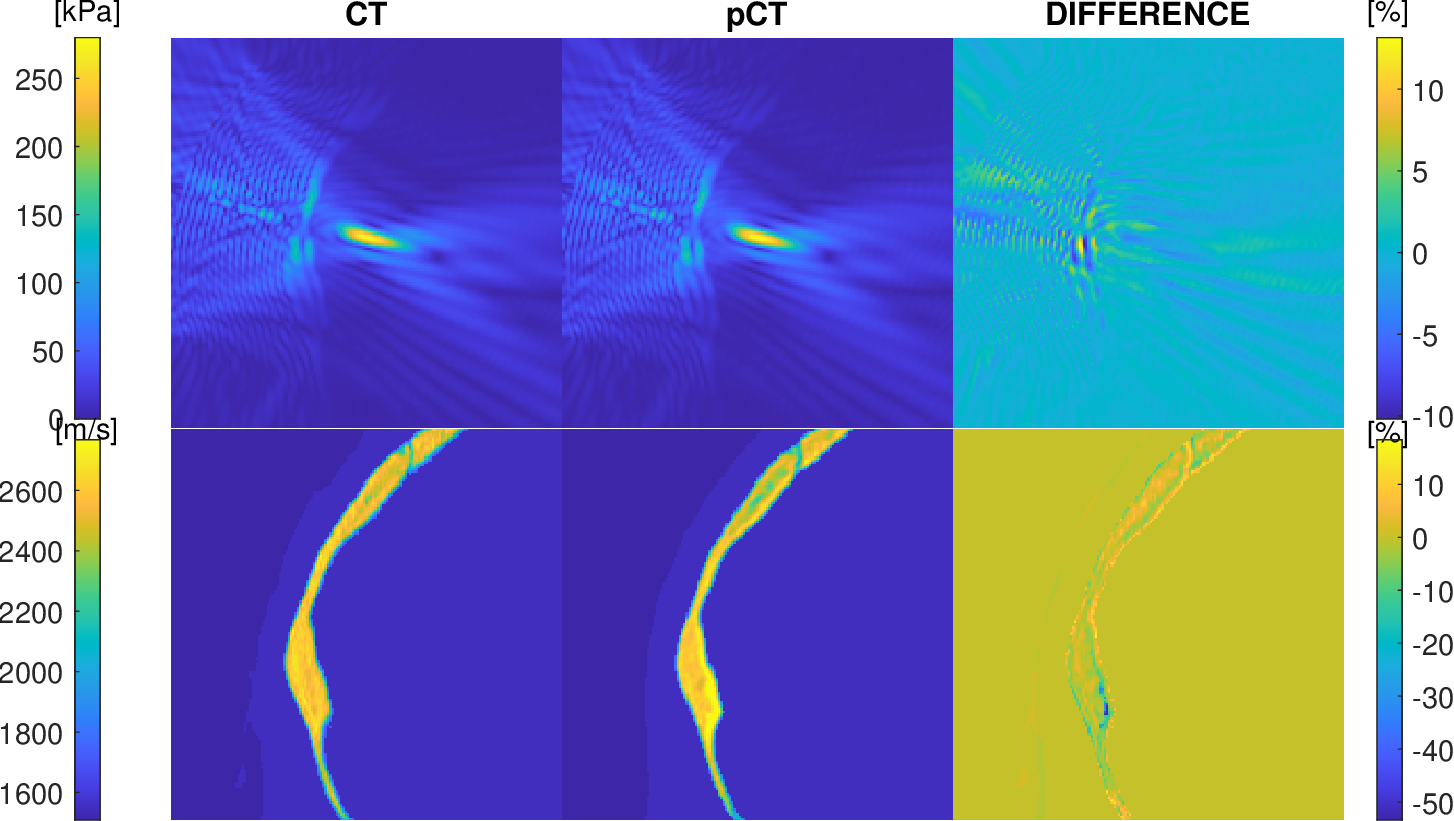}
	    \caption{right visual cortex}
	\end{subfigure}
    \caption{Examples of simulated pressure fields (top) and sound speed maps (bottom) in the four different targets, along with their respective error maps.}
    \label{fig:simulations} 
\end{figure*}

The simulated acoustic fields were evaluated with the metrics described above. 
Examples of the resulting acoustic fields are given in Fig.~\ref{fig:simulations}.

\begin{figure}[h]
    \centering
    \includegraphics[width=0.5\textwidth]{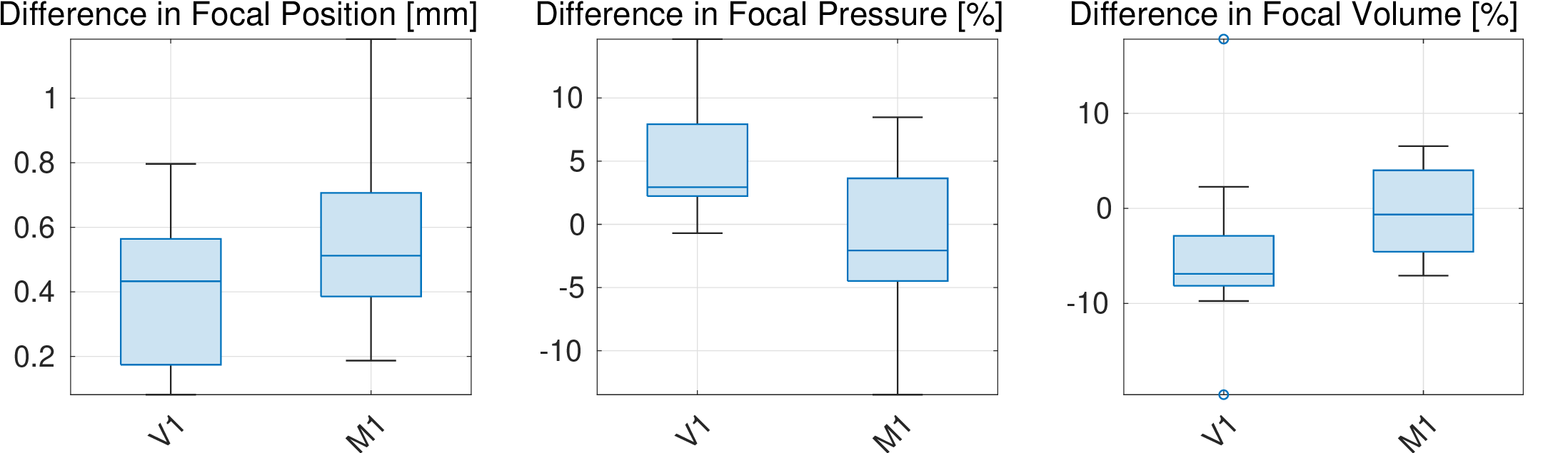}
    \caption{Errors in focal position, pressure and volume, separately for each cortex.}
    \label{fig:acoustic_metrics_boxchart}
\end{figure}

Across all skulls and all targets, the mean difference in focal position was $0.48 \pm 0.25 $ mm, and the mean difference in peak pressure and the mean difference in focal volume were $5.1 \pm 4.0 \%$ and $5.8 \pm 4.4 \% $ respectively. Results for all simulations are summarised in Fig.~\ref{fig:acoustic_metrics_boxchart}. All simulations except one have an error in position smaller than 1 mm, suggesting accurate targeting of the intended brain structures. For the visual cortex, the maximum pressure amplitude tends to be over-predicted in pCT simulations compared to CT simulations, and the focal volume tends to be under-predicted. Results in the visual cortex are more evenly spread.
The above results demonstrate that acoustic simulations based on mapping pCT images from PETRA MRIs can give comparable results to simulations based on ground truth CT, with very low errors in focal position and volume.

Note that the masks used to segment the PETRA image and generate the pCT originate from the PETRA image itself using SPM, since this is designed to work in the absence of a CT. Since in this study CTs were available, this work was reproduced using the CT-derived masks, generated as described in Section \ref{sec:preprocessing}, as the segmentation they provide is more accurate and this allows us to separate mapping errors from segmentation errors. Pseudo-CTs were generated using the CT-derived mask and evaluated using the same methods as before. In this case, the MAE across all subjects is $129 \pm 6$ HU in the head mask, $313 \pm 19$ HU in the skull mask and $64 \pm 1$ HU in the brain mask, while the RMSE is $229 \pm 14$ HU in the head mask, $393 \pm 21$ HU in the skull mask and $82 \pm 2$ HU in the brain mask, just slightly lower to the equivalent numbers in Table~\ref{tab:mae_rmse}. The setup of acoustic simulations was also repeated as before.
Across all skulls and all targets, the mean difference in focal position is $0.45 \pm 0.19 $ mm, and the mean difference in peak pressure and in focal volume are $5.5 \pm 3.9 \%$ and $4.7 \pm 3.6 \% $ respectively. All image and acoustic errors are slightly lower in this case. This suggests that while the imperfect segmentation of the MR with SPM introduces errors, it is a feasible alternative solution in the absence of CT. For reference, the average Dice similarity coefficient across subjects is $0.9706 \pm 0.0044$ in the head mask and $0.8467 \pm 0.0138$ in the skull mask.
\\


Note that the pCTs used rely on the calibration CT to create the acoustic values used in the acoustic simulations. To convert directly from normalised PETRA to density ($\rho$) including the adjustment provided by the calibration CT, the following affine relationship can be used:
\begin{equation}\label{eq:petratoctandcal}
\rho = -1076.1 \cdot PETRA + 2238.7
\end{equation}

So far, it has been demonstrated that this method can generate pseudo-CTs that are very similar to real CTs of in-vivo data, both in terms of raw Hounsfield Units and in the acoustic fields they generate. However, since ex-vivo skulls are commonly used for acoustic field studies and validation, it is of interest to apply this method to ex-vivo data. This way, this approach is validated using measurements to provide the ground truth pressure for comparison with simulations performed using pCTs. Next, the same process is applied to ex-vivo skulls and the results are compared with experimental data.


%% file: 4ExVivoMethods.tex
\section{Experimental validation}\label{sec:2}

The main purpose of using MRI scans instead of CT images for acoustic field simulations is to implement pCT in practice. Therefore, it is reasonable to evaluate not only the similarity between field simulation results obtained using CT and pCT, but also the accuracy of the simulations relative to experimental data. The difference in accuracy between transcranial ultrasound field simulations using a skull model based on pCT versus true CT will provide insights into the feasibility of using pCT models in clinical practice. It is also important to confirm that the accuracy of acoustic calculations with pCT remains consistent over a wide range of frequencies commonly used in TUS, where changes in sound speed due to dispersion and attenuation at different frequencies can affect the results.

\subsection{Dataset and Conversion}\label{sec:exvivodata}
\begin{figure*}[ht!]
    \centering   
    \includegraphics[width=1\textwidth]{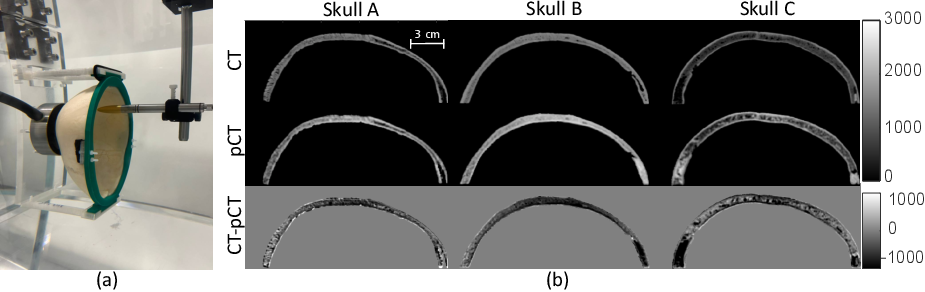}
    \caption{(a) An ultrasound field scanning procedure inside an ex-vivo skull sample inserted into a 3D-printed holder; (b) HU maps of skull specimens, obtained from CT and pCT, and their difference.}
    \label{fig:skulls}
\end{figure*}

In compliance with the UK Human Tissue Act, three human skulls were acquired from the Anatomy Department at King's College London under a material transfer agreement. These skulls, identified as A, B, C are over 50 years old and were cleaned and stored in a dry state. The skulls show variations in characteristics such as porosity, surface smoothness, and overall shape. For this study, only the calvaria were used (as shown in Fig.~\ref{fig:skulls} (a)), which had previously been separated from the lower portions of the skulls. Prior to imaging and experimentation, the skulls were immersed in deionised water and degassed at -400 mbar for 48 hours.

To map the acoustic properties of the skulls, CT scans were conducted using a Siemens SOMATOM Force CT scanner. The scans were performed under the 'Head Spiral' protocol with a tube voltage of 120 kVp, exposure time of 1000 ms, X-ray tube current of 137 mA, and an exposure of 228 mAs, using a flat filter. The images were reconstructed with the $Hr69h\backslash3$ convolution kernel, which is optimised for bone imaging. The pixel spacing was 0.47 x 0.47 mm in-plane, and the slice thickness was 0.5 mm.  A calibration CT was acquired by scanning an electron density phantom (model 062M, CIRS, Norfolk, VA, USA), containing inserts with densities ranging from 1000 to 2200 kg/m$^3$, using the same protocol on the same scanner. Note that these scans are high-dose and were used in the simulations explained in Section~\ref{sec:exvivosims}.

To acquire the relationship from PETRA to CT, low-dose CT and PETRA MR images were acquired with the parameters described in Table~\ref{tab:acquisition_params}. Examples of these images are shown in Fig.~\ref{fig:skulls} (b).
The scans were processed in a similar method as described in Section \ref{sec:pctMethods}. In this case, registration was performed manually in Insight Toolkit \cite{itk}. Metal pin and other artefacts were removed from the scans within Matlab. 
Pseudo-CTs were generated similarly to before; voxels in background/air were assigned to -1000 HU, and voxels in the bone mask were assigned according to the following equation:
\begin{equation}\label{eq:petratoctexvivo}
CT = -2815.1 \cdot MRI + 2779.4.
\end{equation}

Note that this is different from Eq.\ref{eq:petratoctlinear} and it is not possible to apply the mapping obtained from in-vivo data to ex-vivo data. This is because this method depends on soft tissue normalisation using the equivalent histogram peak. In ex-vivo data, there is no soft tissue and the peak in the histogram actually corresponds to the water surrounding the skull. Since water has higher proton density than brain, it appears brighter in an MRI, and the peak has a higher value, leading to different results with the normalisation.

\subsection{Experimental setup and procedure}

The experimental procedure is described in detail in our previous work \citep{krokhmal2025}. In summary, three focused ultrasound transducers were used to generate acoustic fields. Measurements were taken at all four frequencies: 270, 500, 750 kHz and 1 MHz. The H115 model, which consists of a central bowl and an outer ring, was operated at 270 kHz using a 2-channel TPO system from Sonic Concepts (Bothell, WA, USA) connected via an electrical impedance matching network. Additionally, two single-element transducers were used: the H104 transducer, operating at 500 kHz, and the H101 transducer, operating at 750 kHz and 1 MHz. These were controlled by an arbitrary waveform generator (33500B, Keysight, Berkshire, U.K.) and an E\&I 1020L RF amplifier (Electronics and Innovation Ltd., Rochester, NY, USA) through respective electrical impedance matching networks. All transducers featured piezoelectric concave elements with an aperture radius of 32 mm and a curvature radius of 63.2 mm. 

Measurements of the transducers and the acoustic fields within the skulls were performed using a hydrophone positioned in an automated scanning tank filled with deionised water. The measurements employed three hydrophones with submersible preamplifiers: two 0.2 mm PVDF needle hydrophones (Precision Acoustics, Dorchester, U.K.) and a 0.2 mm capsule-type hydrophone (HGL0200, Onda, Sunnyvale, CA, USA). To reduce reflections from the hydrophone holder, the surrounding area was shielded with sound-absorbing rubber. The water temperature was maintained at 20$\degree$C, with fluctuations within $\pm 2 \degree$C during the experiments. Waveforms were captured, digitised, and saved using a digital phosphor oscilloscope (DPO5034B, Tektronix U.K. Ltd., Berkshire, U.K.), controlled by the scanning tank software, with a sampling rate of 125 MHz and 32 averages. The signal amplitude fluctuated due to the interference of the direct wave and multiple reflections within the skull and between the skull and the transducer. At least 10 wave periods after ring up and arrival of all waves were recorded. Signal acquisition was timed to cover the signal range with stabilised amplitude before reflections from the transducer arrived at the measurement location. 

To accurately determine the positioning of the skull specimen in relation to the transducer, as shown in Fig.~\ref{fig:skulls} (a), 3D-printed plastic holders were fabricated. Each skull was fixed in position with its top point placed 10 mm from the transducer's surface. 
Each transducer was characterised using holography \citep{sapozhnikov2015acoustic} to derive source configurations for simulation objectives. 
Holograms were recorded in a free-field environment under quasi-continuous wave (qCW) settings using a sinusoidal pulse of 60 cycles at the target frequency, with a pulse repetition interval of 10 ms. For the minimum frequency, set at 270 kHz, the TPO was activated in a quasi-continuous mode, where signals persisted for 60 $\mu s$ with a 10 ms interval. The same signals were used in experiments with skull specimens. 
Acoustic pressure line measurements for each transducer were performed immediately before mounting the skull, facilitating normalisation of the source amplitude for computational simulations. Further measurements were obtained within the skull cavity with the hydrophone scanned over a plane perpendicular to the beam axis at a distance of 55 mm from the transducer. The scanning area was 60x60 mm, centred at the beam axis in the absence of the skull, with a step size of 0.5~mm. 

The angular spectrum approach (ASA) \citep{schafer1989transducer} was used to reconstruct the acoustic pressure field starting at the transducer surface and reaching an axial distance of 100 mm. The peak positive pressure was determined at each point, and compared with the simulated field.

\subsection{Simulation procedure}\label{sec:exvivosims}

To simulate ultrasound propagation through cranial bone, version 1.4 of the open-source k-Wave toolbox was used. It was chosen as a simulation tool to allow the use of a holographic source plane. The detailed description of the simulation procedure can be found in our earlier publication \citep{krokhmal2025}.To ensure comprehensive coverage, the grid was designed to extend a minimum of 100 mm along the z-axis and fully include the source plane. The spatial resolution was configured to achieve at least 8 points per wavelength, with steps ranging from 0.15 mm at a frequency of 1 MHz to 0.35 mm at 270 kHz. The boundaries of the computational domain were encased in a 20-point thick perfectly matched layer (PML) to mitigate any reflective phenomena. Executions of the simulations occurred on a high-performance computing server outfitted with NVIDIA Tesla P40 GPUs, featuring a memory capacity of 24 GB and a bandwidth of 346 GB/s. Depending on the volume of the domain, simulations might require up to 3 hours to finalise.

The volume surrounding the skull was defined as water at 20$\degree$C with its speed of sound, density and attenuation coefficient as 1483 m/s, 998.2 kg/m$^3$,  $0.0022\cdot f^2$ dB/cm, correspondingly. To calculate the sound speed ($c$) and bone density ($\rho$), cranial tissue densities were determined from HU obtained from CT and pCT images. Density was converted from HU based on the calibration CT scan described in Section \ref{sec:exvivodata}, applicable to both CT and pCT images. For the pCT images, calibration was taken from the CT scanner used in the training process. Mapping of skull acoustic properties was performed according to Section~\ref{sec:acoustic_simulation}.

The simulations utilised an equivalent mass source hologram derived from the measurements acquired for characterisation of each source. 
From these holograms, the k-Wave function \texttt{calculateMassSourceCW} was used to calculate the source \cite{treeby2018equivalent}. The duration of the emitted signal was sufficient to account for reflections within the skull. The amplitude and phase of the spectrum were determined at the specified frequency. To generate a time-varying source in the computational model, the \texttt{createCWSignal} function from the k-Wave toolbox was employed. This function produces a continuous wave signal based on a 2D matrix representing the ultrasound signal amplitude and phase at each point of the source, derived from the measured source holograms. For experimental validation, the simulations provided a 3D field of peak positive pressure at each calculation point. 

\subsection{Metrics for comparing calculations and experiment}\label{sec:metrics}

Simulations were initially performed using the equivalent source hologram in free field conditions to verify that the sources accurately predicted the measured fields. The calculated pressure distribution along the beam axis was compared with experimentally measured data in the free field. A correction coefficient, representing the ratio of maximum amplitudes, was used to adjust the source amplitude for skull simulations. Specifically, the source amplitude in k-Wave for propagation through the skull was scaled by the ratio $p^{exp}_{water}/p^{sim}_{water}$ for each case. This normalisation technique reduced the discrepancies arising from varying hydrophone sensitivity and source output during hologram acquisition and skull measurement. Generally, these variations are rectified by adjusting the drive voltage, yet this method also accounts for any alterations in hydrophone sensitivity.
The focal position, pressure and volume were evaluated, using the same metrics as defined in Section \ref{sec:acoustic_metrics}.
In this case, the subscripts 1 and 2 correspond to simulated fields and experimental measurements respectively.


%% file: 4ExVivoResults.tex
\subsection{Evaluation of pressure fields}\label{sec:3}

\begin{figure*}[t]
    \centering
    \includegraphics[width=\textwidth]{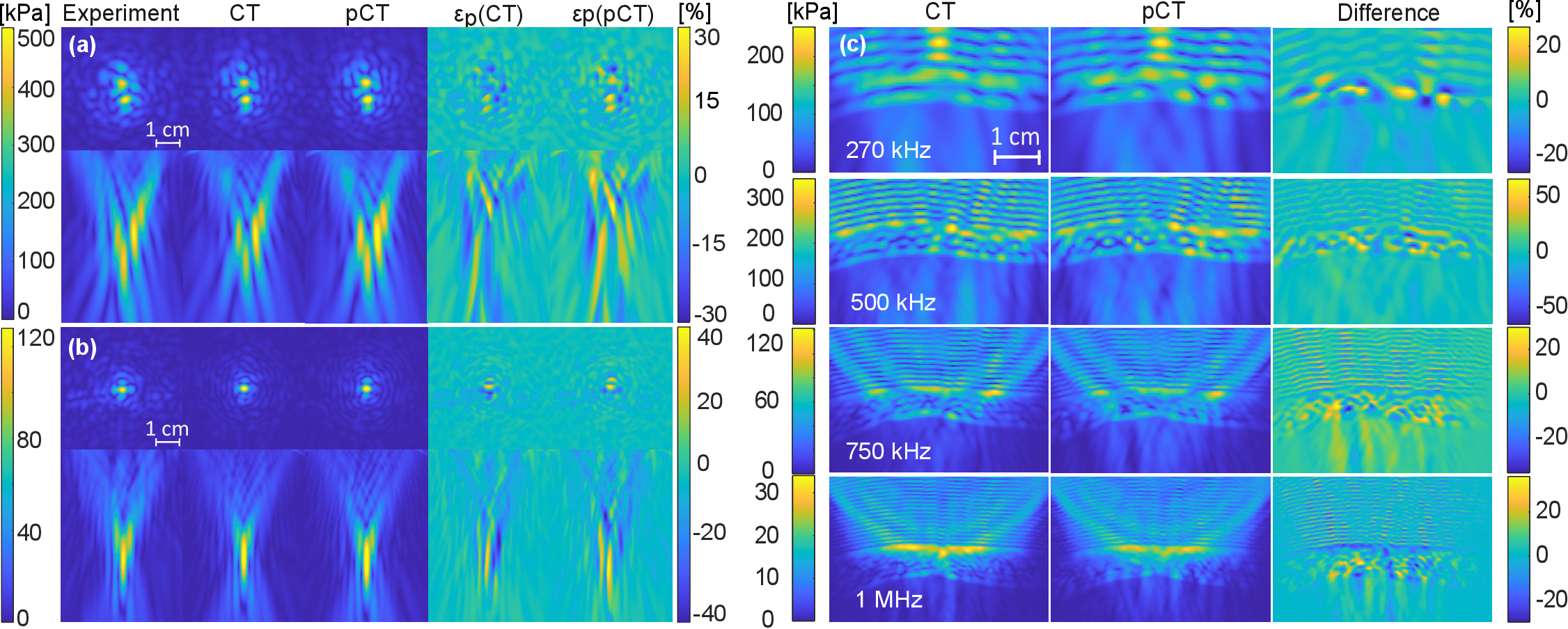}
    \caption{(a) Pressure fields, measured experimentally and obtained from simulations using CT and pCT, and their respective error maps, for (a) 500 kHz and skull sample B and (b) 750 kHz for skull sample C, (c) pressure fields within the skull C at 270 kHz - 1 MHz, obtained from CT and pCT, and their difference.}
    \label{fig:fields}
\end{figure*}

The use of pCT in acoustic simulations did not result in significant distortions of the transcranial ultrasound field compared to conventional CT. In all the cases studied, across the entire frequency range, the spatial distribution of peak pressure for both CT and pCT accurately reproduced the experimental acoustic field structure, as well as the position and shape of the focal spot. An analysis was conducted to evaluate the deviations of numerical simulation results from experimental data, focusing on focal position, peak pressure amplitude, and focal zone volume. Table~\ref{tab:errors_exvivo} below presents the average errors of $r$, $\varepsilon_p$ and $\varepsilon_V$ for both CT and pCT cases relative to experimental results, and comparing to each other. In general, simulations with both pCT and CT reproduce the measured field distribution well. Both have similar errors in focal position, pressure amplitude and focal volume, coinciding within the standard deviation. Across all skulls and frequencies, the mean differences in focal position, peak pressure and focal volume compared to experiments are 1.5±1.2 mm, 17.1±14.6\% and 32.5±22.5\% for CT and 1.6±1.4 mm, 18.2±17.1\% and 19.2±15.9\% for pCT. In calculating the average error for the focal position $r$, two outliers were removed: skull B, characterised by an irregular shape, introduces significant aberrations, dividing the focal spot into distinct segments at frequencies of 270 and 500 kHz, as demonstrated in Fig.~\ref{fig:fields} (a). In this case, although the overarching field structure is accurately reproduced, the location of the peak pressure may vary between spots, leading to significant errors in determining the focal point. 

\begin{table}[h] 
    \begin{tabular}{l| c| c| c}
      &CT \& exp &pCT \& exp & pCT \& CT \\
    \hline
    $r$, mm & 1.5± 1.2   & 1.6±1.4 & 0.8±0.6        \\
     $\varepsilon_p$, \%   & 17.1±14.6   & 18.2±17.1  & 7.9±4.9  \\
     $\varepsilon_V$, \%  & 32.5±22.5 & 19.2±15.9 & 8.5±5.5
    \end{tabular}
    \caption{Average and standard deviation of $r$, $\varepsilon_p$ and $\varepsilon_V$, across 3 subjects and 4 frequencies. Acoustics fields were measured experimentally and compared against the simulations based on CT and pCT images, and also a comparison between CT and pCT based simulations.}
    \label{tab:errors_exvivo}
\end{table}

\begin{figure*}[t]%
    \centering
    \includegraphics[width=\textwidth]{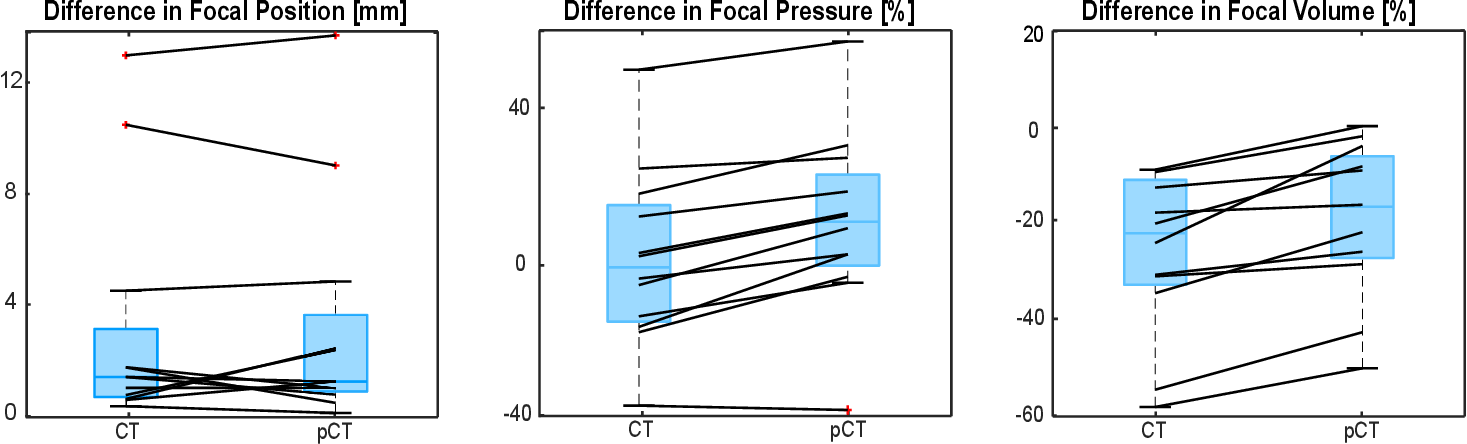}
    \caption{Boxcharts of errors in focal position, pressure and volume, over 3 ex-vivo skull samples and 4 frequencies. Black lines connect individual cases. Outliers are displayed using +.}
    \label{fig:all_values}
\end{figure*}

Fig.~\ref{fig:all_values} compares the individual errors and their statistical distribution in focal position, peak pressure, and focal volume between conventional CT and pCT for transcranial ultrasound simulations across three ex-vivo skull samples and four frequencies. The results demonstrate that pCT offers comparable accuracy relative to CT, with reduced variability in most parameters. 
For focal position, the median error is lower for pCT (1.3 mm) than CT (1.5 mm), indicating a more consistent reproduction of focal positions. However, the interquartile range (IQR) is greater for pCT (2.8 mm vs 2.5 mm for CT), reflecting increased variability. Outliers correspond to simulations for skull B at 270 and 500 kHz, as mentioned earlier. In terms of focal pressure, pCT shows median error greater than for CT (-1.5\% for CT vs 10\% for pCT), suggesting that pCT overestimates experimental pressure values on average. The trend lines connecting individual cases also reveal that pCT consistently maintains higher pressure amplitude. For focal volume, pCT outperforms CT, with a median difference closer to zero (-22.5\% for CT vs -17.5\% for pCT) and a comparable IQR (21.8\% for CT vs 21.1\% for pCT), indicating improved accuracy.

The analysis of difference between CT and pCT simulations across all the ex-vivo skulls demonstrates slightly higher error comparing to in-vivo skulls: 0.8±0.6 mm for focal position, 7.9±4.9\% for peak pressure and 8.5±5.5\% for focal volume. Nevertheless, simulations using pCT images still provide high level of similarity of the pressure field comparing to simulations based on CT and experiment, replicating the main and additional peaks, and the hot spots within the skull, as shown in Fig.~\ref{fig:fields} (a, b).

In treatments using long-pulse duration, there is an increased risk of creating high pressure regions in the skull spots due to the rise of reflections effects.
A comparative analysis of pressure fields inside ex-vivo skulls obtained through simulations using CT and pCT was performed. In all cases the difference between peak pressure amplitude of hot spots for CT and pCT never exceeded 16\% with no noticeable trend on frequency. The field structure inside skull matched well, with hot spots located in the same areas, as shown in Fig.~\ref{fig:fields} (b) for the skull C. 

Overall, the differences between simulations in CT and pCT are similar, and the field structure doesn't differ, which indicates that pCTs can act as an alternative to CTs.



%% file: 5Conclusions.tex
\section{Discussion}
It is demonstrated specialised MR sequences can be used to convert to the equivalent pCT, with good accuracy. The generated pCTs have similar mean absolute errors to relevant literature, which suggests that errors of 300 HU are common, but can go down to 100 HU especially with machine learning methods. These values are still low, considering that the average value in the skull is 1574 HU.

The pCTs were used to create acoustic simulations and they were compared to those generated from CT. For in vivo-data, the results show that pCTs can create very similar acoustic maps and can target the visual and motor cortices accurately. The errors in the focal position, volume and pressure are very low, which supports that we can accurately target brain structures with PETRA generated pCTs. 

Simulations ran in a water medium showed the focal volume to be 3.4 mm wide and 17.7 mm long. Taking into account that the mean difference in focal position is $0.45 \pm 0.19 $ mm, it means that the intended structures can be accurately targeted. 
For context, in the motor cortex, the smallest relevant target area for stimulation could reasonably be considered the cortical representation of a single hand muscle. A recent transcranial magnetic stimulation (TMS) study suggests that the functional area of the first dorsal interosseous (FDI), a muscle commonly targeted in motor studies, is approximately 26 mm$^2$ \cite{reijonen2020spatial}. In the primary visual cortex, the mean distance between approximate centres of TMS-distinct visual regions (V1 and V2d) is 11 mm \cite{salminen2012selective}, suggesting a similar necessary resolution to M1. The errors in the focal position using the pCT images are significantly less.

The simulations using ex-vivo are based on high-dose scans for real CT and low-dose scans for pseudo-CTs. While that may account for some of the differences, the range of errors shows that this method can be applied on either high-dose or low-dose CTs.

One of the main findings it that the simulations using CT or pCT result in very small errors in all three metrics: focal position, peak pressure and focal volume. This is demonstrated in both in-vivo and ex-vivo cases. 

Another key result is the strong agreement in focal position between pCT and CT simulations compared to experimental data. The average error of approximately 1.5 mm is sufficient to ensure accurate targeting for most clinical applications \cite{robertson2017accurate}.
This limitation should be considered when planning ultrasound interventions at frequencies below 500 kHz.

However, it is evident from the results that both the CT and pCT models exhibit a higher uncertainty in peak pressure amplitude compared to experimental data. This discrepancy arises due to differences between the model's attenuation coefficient and the actual attenuation observed in skull samples.


The errors in focal volume can be attributed to strong aberrations at high frequencies. In particular, skull B has a highly non-uniform shape and the focal volume splits into distinct segments. However, the observed reduction in focal volume errors when using pCT compared to CT suggests that the HU-to-speed-of-sound conversion derived from PETRA data aligns more closely with experimental outcomes. This finding may indicate that pCT offers improved modelling accuracy for specific cases.

Overall, while the accuracy of pressure amplitude simulations depends on the precision of attenuation modelling in the skull, the differences between CT and pseudo-CT are minimal. This indicates that pseudo-CT can reliably be used for treatment planning. 


\section{Summary and conclusion}

In summary, it is demonstrated that an affine mapping method can be used to translate from PETRA scans to pseudo-CT images.
Paired sets of these images were used to acquire this mapping from normalised PETRA values to CT values, given in Eq.~\ref{eq:petratoctlinear} for in-vivo data and Eq.~\ref{eq:petratoctexvivo} for ex-vivo data.
An open repository with code \cite{petra_repo} is provided to generate pseudo-CTs, as well as the exact set of PETRA acquisition parameters needed. 

Acoustic simulations using CTs and pCTs of healthy human participants showed that they generate very similar acoustic fields: errors in focal position, peak pressure, and focal volume were 0.48±0.25 mm, 5.1±4.0\%  and 5.8±4.4\%, respectively.
Experimental validation of the ex-vivo simulations revealed that there was no noticeable rise in average errors when comparing pCT to CT.
Pseudo-CTs consistently demonstrated good agreement in focal position and pressure amplitude, meeting clinical requirements for accurate targeting, while errors in peak amplitude were higher. Overall, pCTs provide comparable accuracy to CTs, supporting their integration into safer and more accessible TUS workflows and usage in treatment planning as an alternative to real CTs.